\documentclass[12pt]{article}

\usepackage{graphicx}
\begin{document}

\begin{center}
{\Large QCD at non-zero temperature: bulk properties and heavy quarks }\\[3mm]
P\'eter Petreczky\\
Physics Department
Brookhaven National Laboratory
Upton, NY, 11973, USA\\[3mm]
\end{center}

\noindent {\bf Abstract:} I review recent progress in lattice QCD at non-zero temperature
whith emphasis on the calculations of equation of state and the properties of heavy quar anti-quark
pairs at high temperatures. I also briefly discuss the deconfinement and chiral symmetry restoring
aspects of the QCD transition at finite temperature

\section{Introduction}
It is expected that strongly interacting matter undergoes a transition in some temperature
interval from hadron gas to deconfined state also called the quark gluon plasma (QGP) \cite{gros81}. 
Creating deconfined medium in a laboratory is 
the subject of the large experimental program at RHIC 
\cite{nagle}
and is going to be the goal of the
future heavy-ion program at LHC \cite{salgado}.
Attempts to study QCD thermodynamics on the lattice go back to the early 80's when lattice
calculations in $SU(2)$ gauge theory provided the first rigorous theoretical evidence for
deconfinement \cite{kuti81,mclerran81,engels81}. The problem of calculating thermodynamic observables in pure
gluonic theory was solved in 1996 \cite{boyd}, while calculation involving dynamical quarks
were limited to large quark masses and had no control over discretization errors 
\cite{milc_old_eos,wilson_old,p4_old} (see Refs. \cite{rev1,lp_rev} for reviews).  
Lattice calculations of QCD thermodynamics with
light dynamical quarks remained challenging until recently. During the past 5 years calculations
with light $u,d$ quarks have been performed using improved staggered fermion actions 
\cite{milc_Tc,fodor05,our_Tc,fodor06,milc_eos,our_eos,eos005,hotQCD,fodor09,fodor10,fodor10eos,wwnd10}
(see Refs. \cite{my_qm09, rev2}.) 

To get reliable predictions from lattice QCD the lattice spacing $a$ should be sufficiently small
relative to the typical QCD scale, i.e. $\Lambda_{QCD} a \ll 1$. For staggered fermions, which are used
for calculations at non-zero temperature, discretization
errors go like ${\cal O}( (a \Lambda_{QCD})^2)$ but discretization errors due to flavor symmetry
breaking turn out to be numerically quite large. 
To reduce these errors one has to use improved staggered fermion actions with so-called fat links
\cite{orginos}. At high temperature the dominant discretization errors go like $(a T)^2$ and therefore
could be very large. Thus it is mandatory to use improved discretization schemes, which improve the
quark dispersion relation and eliminate these discretization errors. Lattice fermion actions 
used in numerical calculations typically implement some version of fat links as well as improvement
of quark dispersion relation and are referred to as $p4$, $asqtad$, $HISQ$ and $stout$. In lattice calculations
the temperature is varied by varying the lattice spacing at fixed value of the temporal extent $N_{\tau}$.
The temperature $T$ is related to lattice spacing and temporal extent, $T=1/(N_{\tau} a)$. Therefore
taking the continuum limit corresponds to $N_{\tau} \rightarrow \infty$ at the fixed physical volume.
For the same reason discretization errors in the hadronic phase could be large when the temperature $T$ is small.

In this paper I am going to discuss lattice QCD calculation of the transition temperature,
equation of state as well as different spatial and temporal correlation functions relevant for
phenomenology of heavy quarks.

\section{Equation of State}
\begin{figure}[ht]
\includegraphics[width=7.3cm]{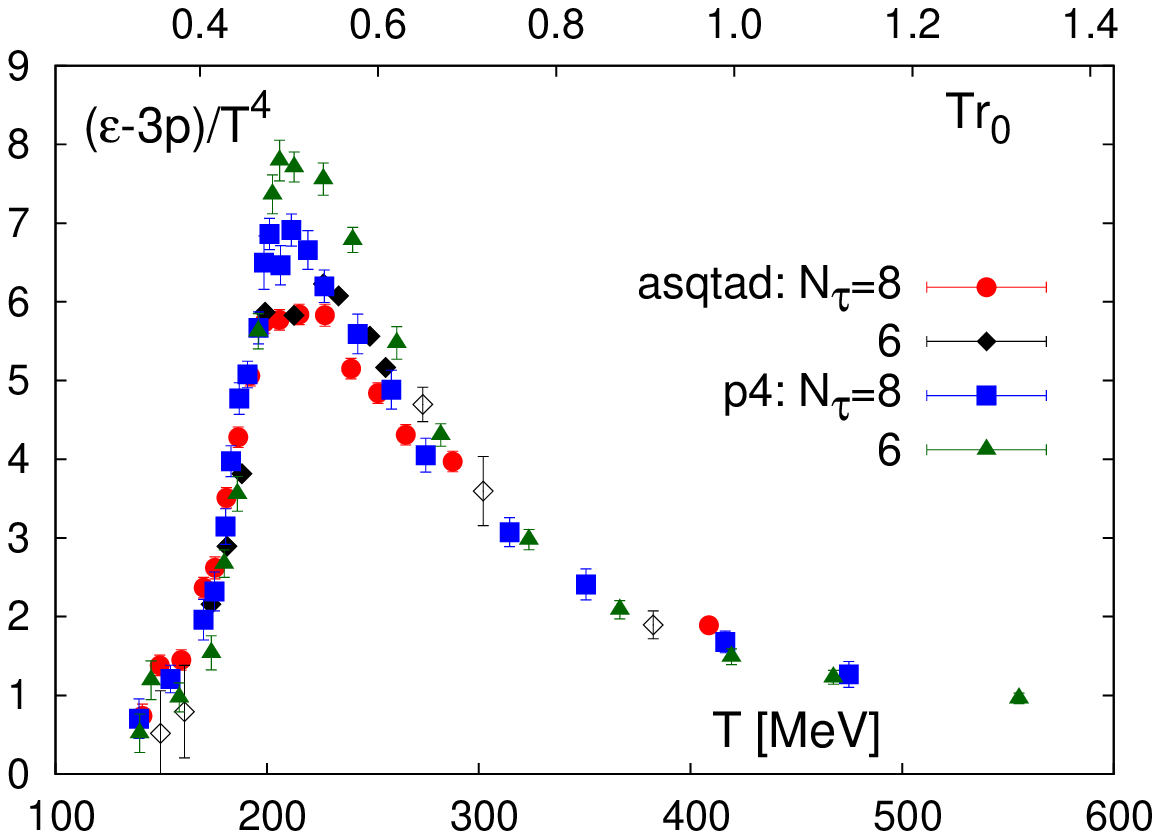} 
\includegraphics[width=7.3cm]{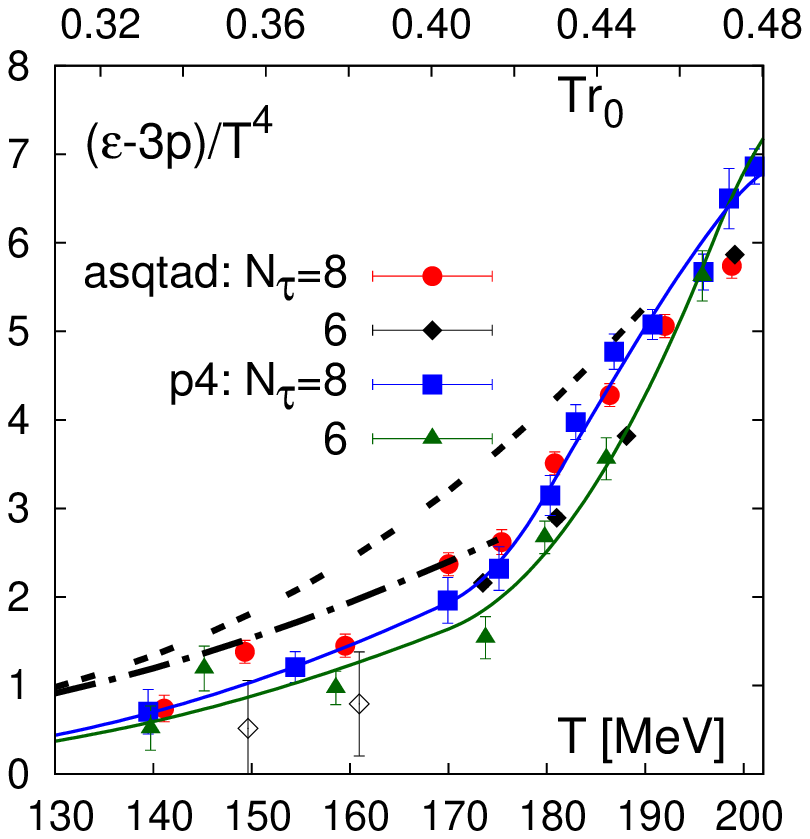}
\vspace*{-0.3cm}
\caption[]{
The interaction measure calculated with $p4$ and $asqtad$ actions in entire temperature range (left)
and at low temperatures (right) from Ref. \cite{hotQCD}. The dashed and dashed-dotted lines are
the prediction of hadron resonance gas (HRG) 
with all resonances included up $2.5$GeV (dashed) and $1.5$GeV (dashed-dotted), respectively.
}
\label{fig:e-3p}
\end{figure}

The equation of state has been calculated with $p4$ and $asqtad$ action on lattices with temporal extent
$N_{\tau}=4,~6$ and $8$ \cite{milc_eos,our_eos,hotQCD}. In these calculations the strange quark mass
was fixed to its physical value, while the light ($u,~d$) quark masses 10 times smaller than
the strange quark mass have been used. These correspond to pion masses of $(220-260)$ MeV. The calculation
of thermodynamic observables proceeds through the calculation of the trace of the energy momentum tensor
$\epsilon -3 p$ also
known as trace anomaly or interaction measure. This is due to the fact that this quantity can be expressed in
terms of expectation values of local gluonic and fermionic operators. The explicit expression for $\epsilon -3 p$
in terms of these operators for $p4$ and $asqtad$ actions can be found in Ref. \cite{hotQCD}.  
Different thermodynamic observables can be obtained from the interaction measure through integration. The
pressure can be written as
\begin{equation}
\displaystyle
\frac{p(T)}{T^4}-\frac{p(T_0)}{T_0^4}=\int_{T_0}^T \frac{dT'}{T'^5} (\epsilon-3 p).
\end{equation}
The lower integration limit $T_0$ is chosen such that the pressure is exponentially small there.
Furthermore, the entropy density can be written as $s=(\epsilon+p)/T$. Since the interaction measure 
is the basic thermodynamic observable in the lattice calculations it is worth discussing its properties
more in detail. In Fig. \ref{fig:e-3p} I show the interaction measure for $p4$ and $asqtad$ actions
for two different lattice spacings corresponding to $N_{\tau}=6$ and $8$. In
the high temperature region, $T>250$ MeV results obtained 
with two different lattice spacings and two different actions
agree quite well with each other. Furthermore, recent calculations with HISQ actions also give results 
for the trace anomaly which are consistent with these \cite{wwnd10}.
Discretization errors are visible in the temperature region, where
$\epsilon -3 p$ is close to its maximum as well as in the low temperature region. At low temperatures the lattice
data have been compared with the hadron resonance gas (HRG). 
As one can see the lattice data fall below the resonance
gas value. This is partly due to the fact that the light quark masses  are still about two times larger than
the physical value as well as to discretization errors. The ${\cal O}( (a \Lambda_{QCD})^2)$ 
discretization errors in the hadron spectrum are suppressed at high temperatures as the lattice spacing $a$ is
small there. Also hadrons are  not the relevant degrees of freedom in this temperature region. But at small
temperatures, where hadrons are the relevant degrees of freedom, these discretization effects are significant.
It turns out, however, that 
the HRG model that takes into account the quark mass dependence and discretization errors in the hadron
spectrum can describe the lattice data quite well \cite{pasi,pasi1}.
The large discretization errors in the low temperature region is the reason 
for the discrepancies with the stout results at temperatures
$T<200$ MeV \cite{fodor10eos}. The lattice results for the trace anomaly obtained with stout action are also
different in the high temperature limit \cite{fodor10eos}. 
Namely, $(\epsilon-3p)/T^4$ calculatied with stout action is about 50\% smaller in the high temperature region.
To resolve this problem calculation at finer lattice
spacing are needed.

\begin{figure}[ht]
\includegraphics[width=7.5cm]{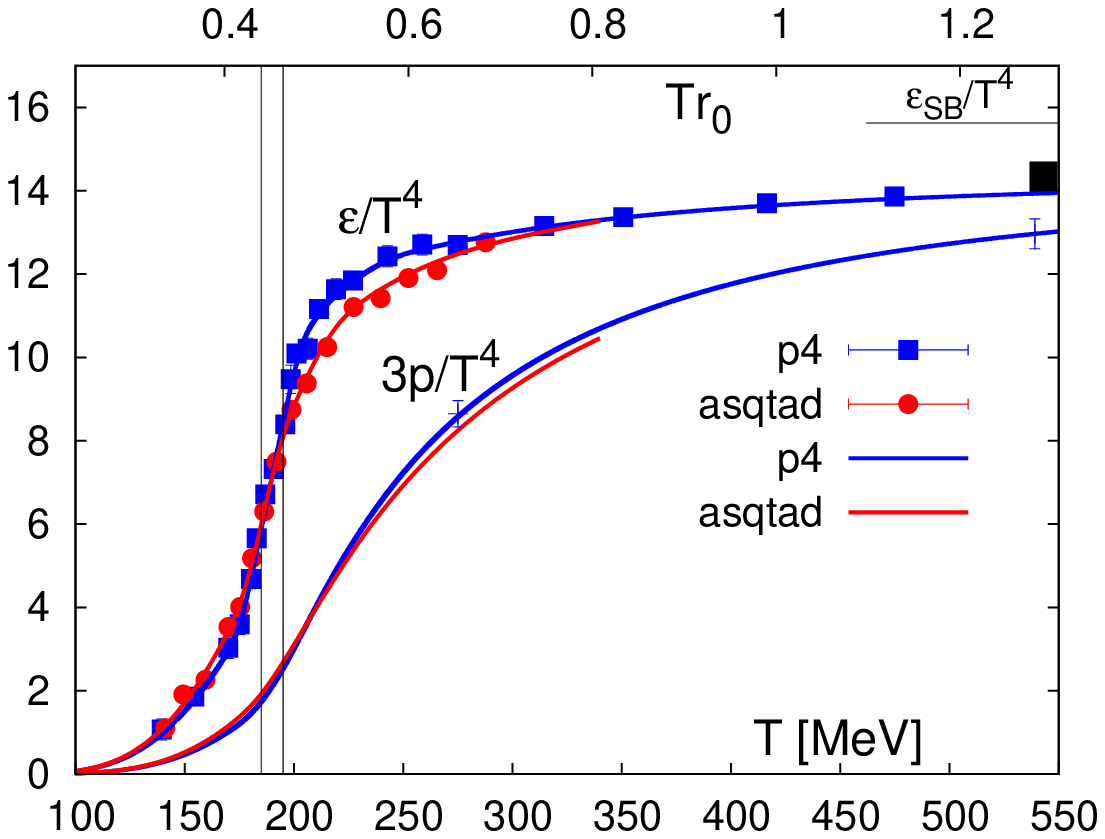}
\includegraphics[width=7.5cm]{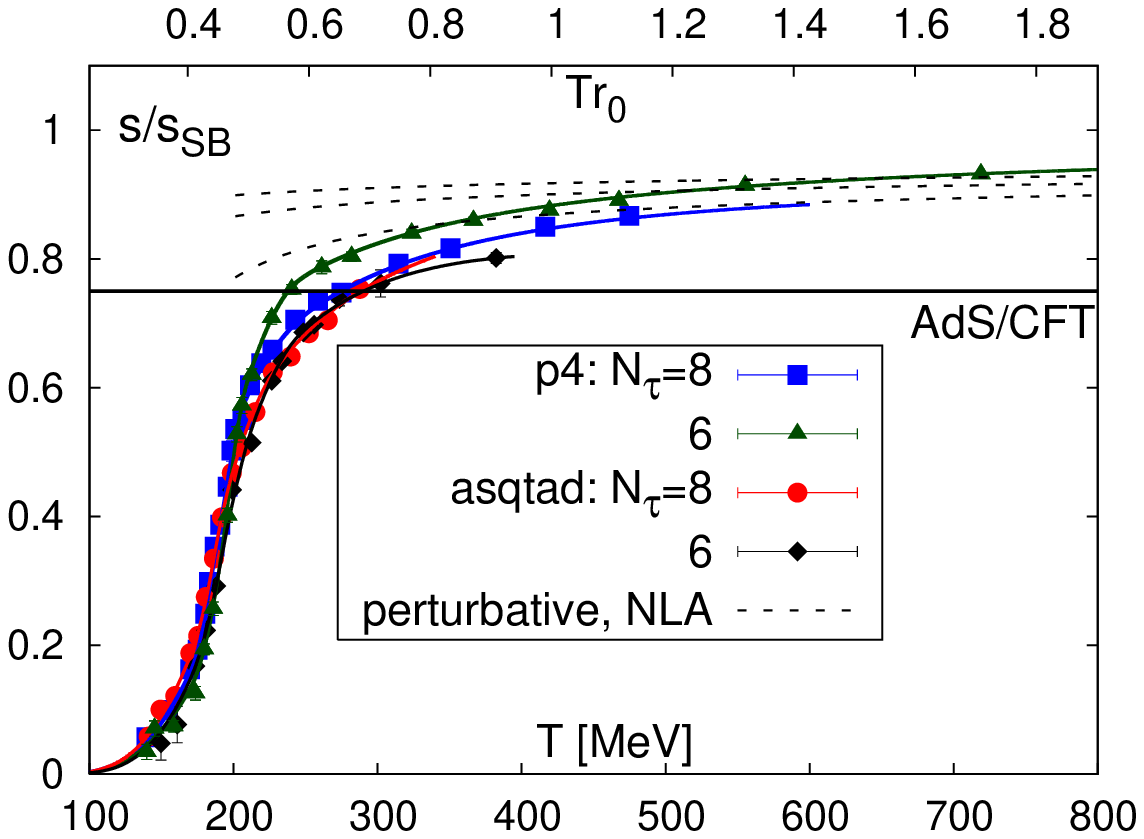}
\vspace*{-0.3cm}
\caption[]{
The energy density and the pressure as function of the temperature (left), and the entropy density divided
by the corresponding ideal gas value (right). The dashed lines in the right panel correspond to the resummed
perturbative calculations while the solid black line is the AdS/CFT result.
}
\label{fig:eos}
\end{figure}

The pressure, the  energy density and the entropy density are shown in Fig. \ref{fig:eos}. The energy density
shows a rapid rise in the temperature region $(185-195)$ MeV and quickly approaches about $90\%$ of the ideal gas
value. The pressure rises less rapidly but at the highest temperature it is also only about $15\%$ below the ideal
gas value. In the previous calculations with the $p4$ action it was found that the pressure and energy density
are below the ideal gas value by about $25\%$ at high temperatures \cite{p4_old}.
Possible reason for this larger deviation could
be the fact that the quark masses used in this calculation were fixed in units of temperature instead 
being tuned to give constant meson masses as lattice spacing is decreased. As discussed
in Ref. \cite{fodor04} this could reduce the pressure by $10-15\%$ at high temperatures. 
In Fig. \ref{fig:eos} I also show the entropy density divided by the corresponding ideal gas value
and compare the results of lattice calculations with resummed perturbative calculation \cite{blaizot,blaizot1}
as well as with the predictions from AdS/CFT correspondence for the strongly coupled regime \cite{gubser98}.
The later is considerably below the lattice results. Note that pressure, energy density and the trace anomaly
have also been recently discussed in the framework of resummed perturbative calculations which seem to
agree with lattice data quite well \cite{mike}.

The differences between the $stout$ action and the $p4$ and $asqtad$ actions for the trace anomaly
translates into the differences in the pressure and the  energy density. 
In particular, the energy density is about 20\% below the ideal gas limit for the $stout$ action. 

\section{Fluctuations of conserved charges}
\begin{figure}[ht]
\includegraphics[width=7cm]{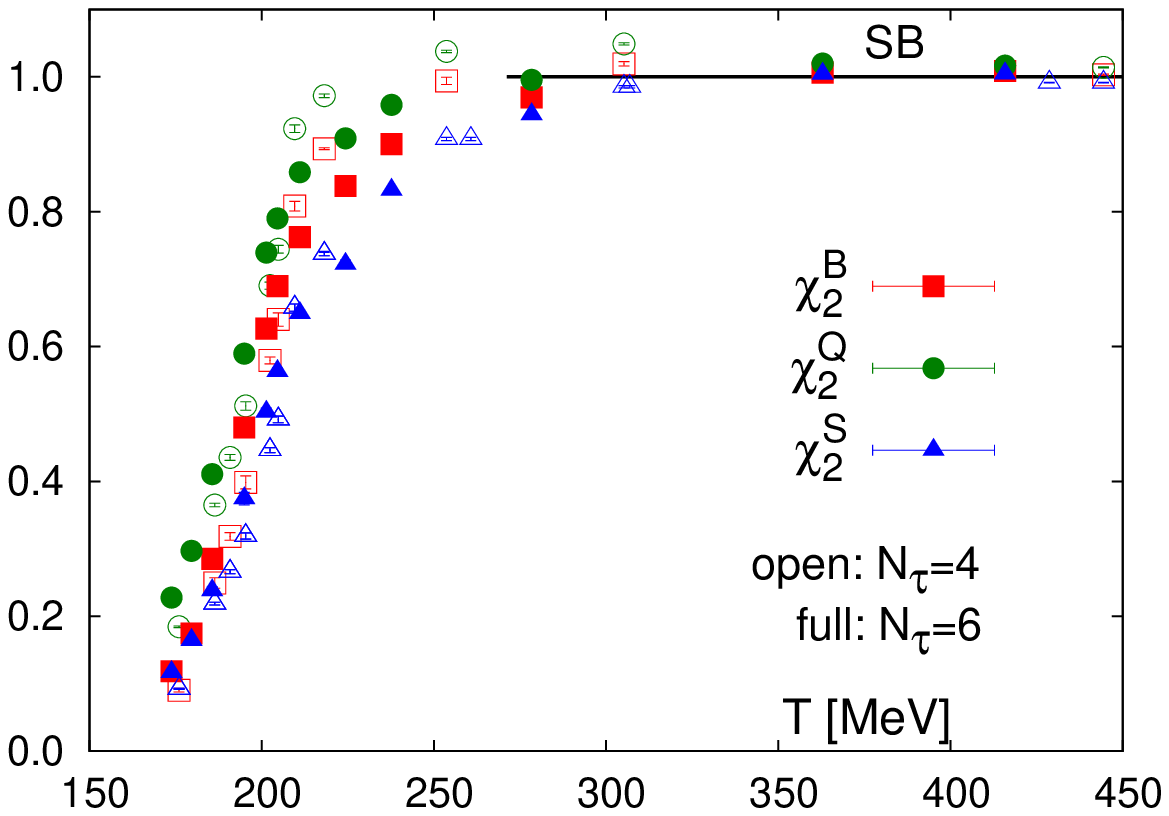}
\includegraphics[width=7cm]{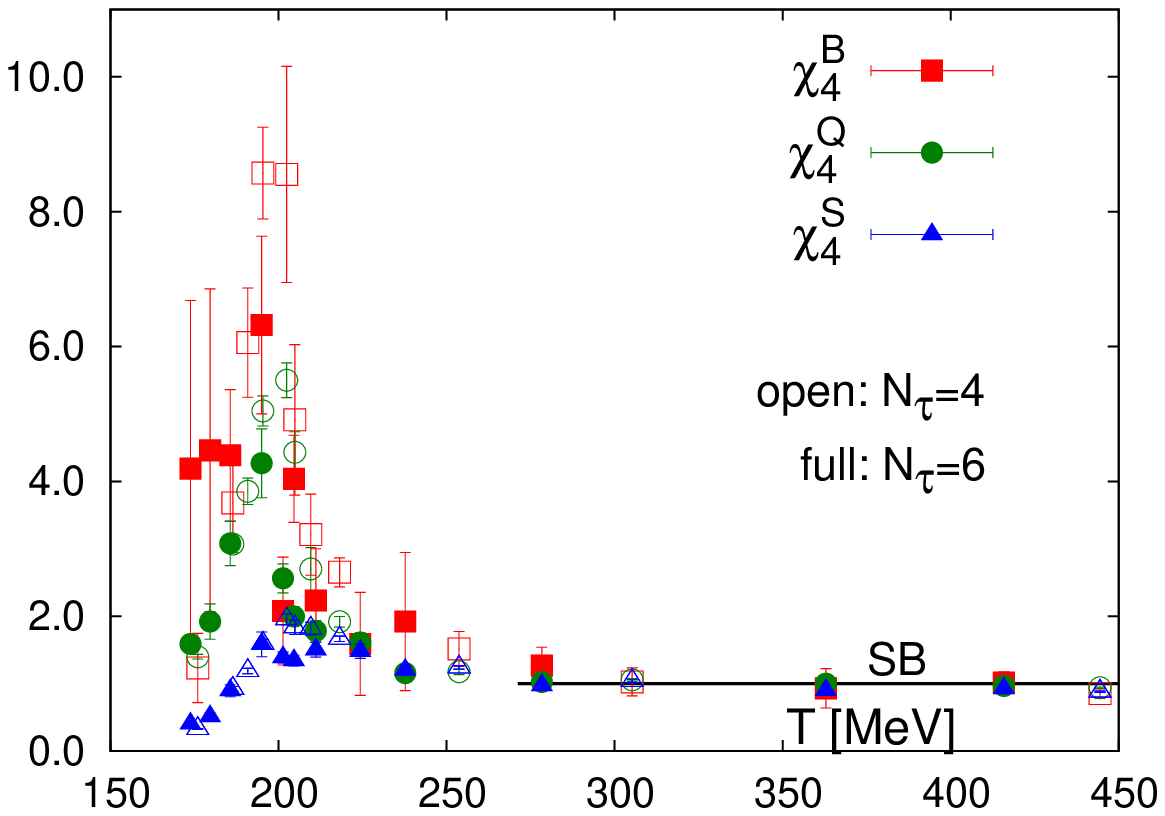}
\vspace*{-0.3cm}
\caption[]{
The quadratic (left) and quartic (right) fluctuations of $B$, $Q$ and $S$
as function of the temperature \cite{our_fluct}. The open symbols correspond to the results
obtained on $N_{\tau}=4$ lattices, while the filled symbols to the ones
obtained on $N_{\tau}=6$ lattices.
}
\label{fig:chi}
\vspace*{-0.2cm}
\end{figure}

The pressure at non-zero chemical potentials can be evaluated using Taylor expansion. The 
Taylor expansion can be set up in terms of quark chemical potentials or in terms of chemical
potentials corresponding to baryon number $B$, electric charge $Q$ and strangeness $S$ of hadrons.
The expansion coefficients in quark chemical potential $\chi_{uds}^{jkl}$ and 
the hadronic ones $\chi_{BQS}^{jkl}$ are related to each other. For a more detailed discussion on this 
see Ref. \cite{our_fluct}.
Taylor expansion can be used to study the physics at non-zero baryon density. However, the expansion
coefficients also give important information about fluctuation of conserved charges, which turn out
to be sensitive probes of deconfinement and chiral symmetry restoration. Fluctuations of conserved charges
have been studied in detail using $asqtad$ and $p4$ actions \cite{milc_Tc,hotQCD,our_fluct}. In Fig. \ref{fig:chi}
I show the quadratic and quartic fluctuations of $B,~Q$ and $S$. Fluctuations are suppressed at low temperatures
because conserved charges are carried by massive hadrons, but rapidly grow in the transition region 
$T=(185-195)$ MeV as consequence of deconfinement. At temperatures $T>300$ MeV fluctuations are reasonably
well described by ideal gas of quarks. Quartic fluctuations exhibit a peak near the transition region, which
at sufficiently small quark masses can be related to the chiral transition and the corresponding critical
exponents (see discussion in Refs. \cite{hotQCD}). As one can see from the figure quadratic fluctuations
show visible cutoff dependence at low temperatures, while the cutoff dependence is small for $T>250$ MeV.

\begin{figure}[ht]
\includegraphics[width=7cm]{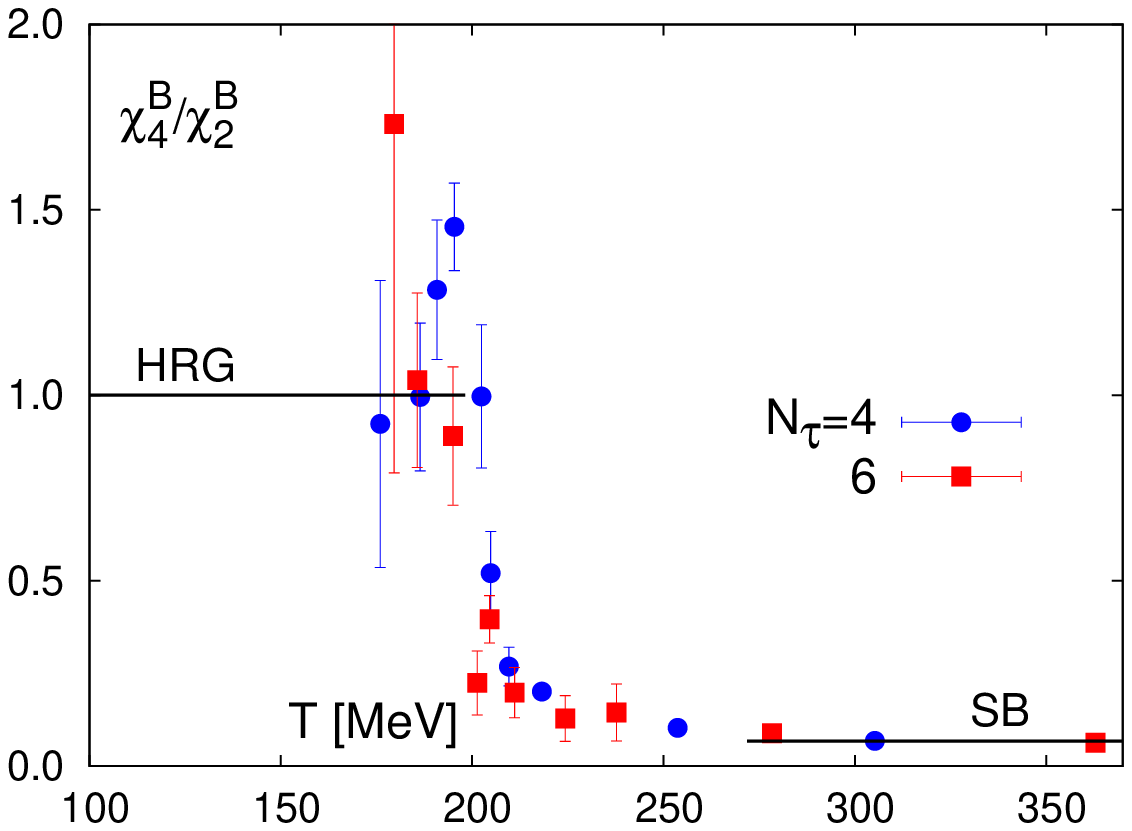}
\includegraphics[width=7cm]{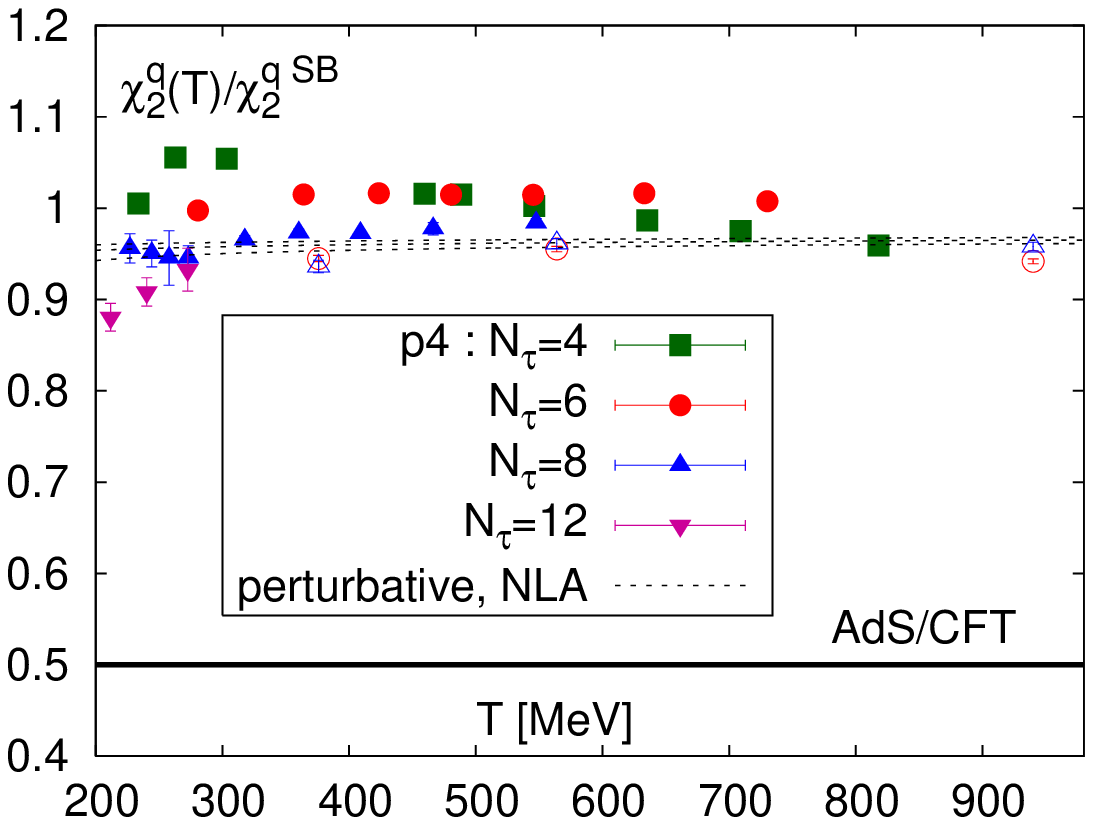}
\vspace*{-0.3cm}
\caption[]{
The Kurtosis of baryon number (left) and the quark susceptibility
compared with resummed perturbative calculations (right). 
The open symbols on the right plot correspond to $asqtad$ results 
for strange quark number susceptibility \cite{milc_Tc}.
The resummed perturbative
results were obtained in next-to-leading approximation (NLA) \cite{rebhan_sewm02}.
}
\label{fig:chi_highT}
\vspace*{-0.5cm}
\end{figure}

The transition from hadronic to quark degrees of freedom can be particularly well seen in the temperature
dependence of the Kurtosis, which is the ratio of quartic to quadratic fluctuations. This quantity can
be also measured experimentally. In Fig. \ref{fig:chi_highT} I show the Kurtosis
of the baryon number as a function of the temperature. At low temperatures it is temperature independent
and is close to unity in agreement with the prediction of the hadron resonance gas model. In the transition
region it sharply drops from the hadron resonance gas value to the value corresponding to an ideal gas
of quarks. Since the fluctuations of conserved charges are so well described by ideal quark gas it is interesting
to compare the lattice results for quark number fluctuations with resummed perturbation theory. 
Quadratic quark number fluctuations are also called quark number susceptibilities and have been calculated
in resummed perturbation theory \cite{blaizot01,rebhan_sewm02,munshi}. The comparison of the resummed perturbative
results with lattice calculations performed with $p4$ action \cite{progress} 
for the quark number susceptibility divided
by the ideal gas value is shown in Fig. \ref{fig:chi_highT}. The figure shows a very good agreement
between the lattice calculations and resummed perturbative results. It is also interesting to compare findings
of the lattice calculations with the results obtained in strongly coupled gauge theories using AdS/CFT 
correspondence. The result from AdS/CFT calculations shown in the figure as the solid black line is significantly
below the lattice results \footnote{the conserved charges considered in these calculations are not exactly the
quark numbers, see discussion in Ref. \cite{my_qm09}.}.
Let me finally note that the discretization errors
in the lattice calculations of the fluctuations are small at high temperatures when calculated with
the $p4$ action. The same is true for $asqtad$ action \cite{milc_Tc,hotQCD}. 

\section{Chiral and deconfinement transition}
\begin{figure}
\includegraphics[width=0.50\textwidth]{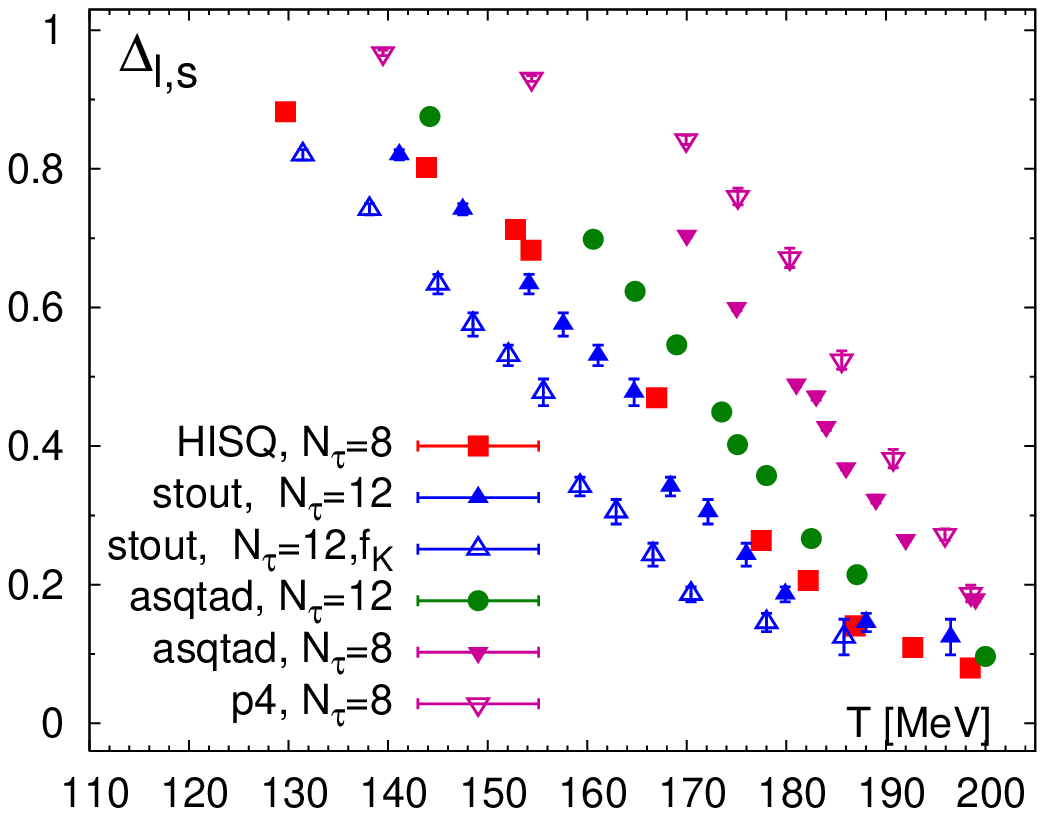}
\includegraphics[width=0.50\textwidth]{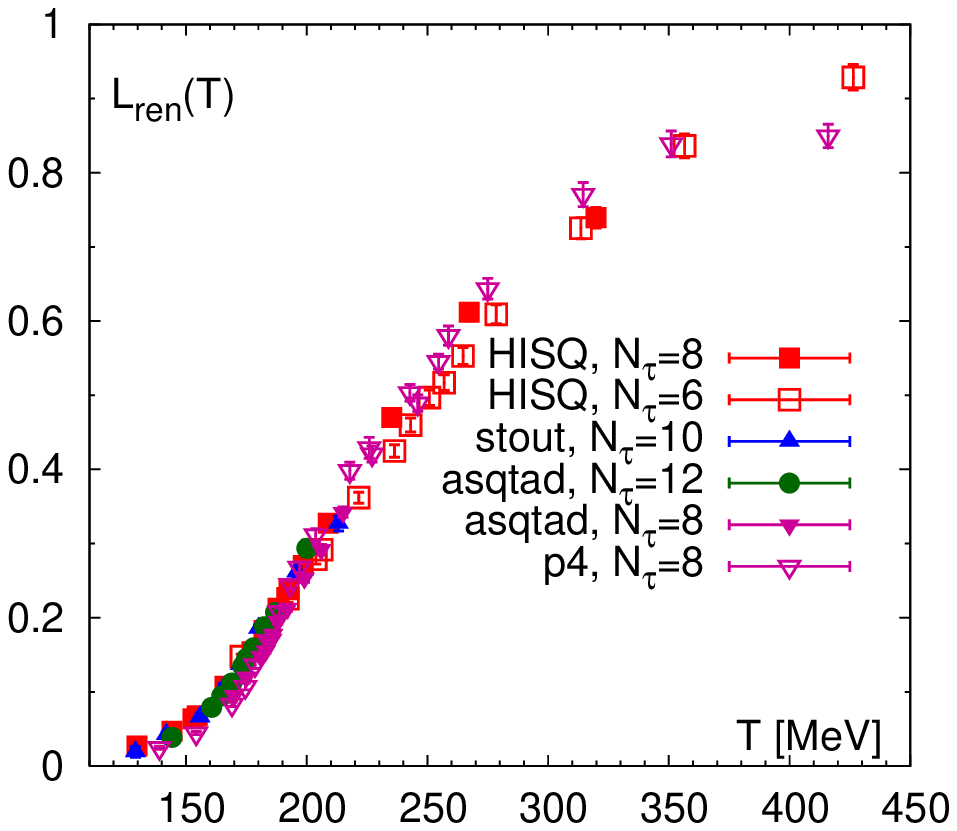}
\caption{The subtracted chiral condensate (left) and the renormalized Polyakov loop (right)
calculated for different actions \cite{eos005,fodor09,wwnd10}.}
\label{fig:order}
\end{figure}

The finite temperature transition in QCD has aspects related to
deconfinement and chiral symmetry restoration. Deconfinement aspects
of the transition are related to color screening and, for infinitely
heavy quarks, also to the center symmetry. The order parameter for
deconfinement is the Polyakov loop which is related to the free
energy of isolated static quark. The breaking of the center symmetry is
signaled by non-zero value of the Polyakov loop.
In the opposite limit of massless quarks
QCD has the chiral symmetry. The quark condensate $\langle \bar \psi \psi \rangle$
is the order parameter for this symmetry. The chiral symmetry is broken in the vacuum
and expected to be restored at high temperatures. In Fig. \ref{fig:order} the renormalized
Polyakov loop and the subtracted chiral condensate are shown. Both quantities show a smooth
change in the transition region which is consistent with the fact that the 
finite temperature transition is an analytic crossover and not a true phase transition \cite{nature}.
It is not clear to what extent the center symmetry is relevant for the physical values of
the quark masses and it is difficult to define a deconfinement transition temperature.
The universal aspects of the chiral transition seem to be relevant for the range of quark masses
which include the physical light quark mass and the corresponding transition temperature can
be defined. Calculations with the p4 action on $N_{\tau}=4$ and $6$ lattices gave an estimate
$T_c=192(4)(7)$MeV for the chiral transition temperature \cite{our_Tc} which is significantly larger than
the value of about $155$MeV obtained for stout action. This is due to large flavor symmetry breaking
for the p4 action. As one can see from Fig. \ref{fig:order} the discrepancies between different
actions are reduced when considering larger $N_{\tau}$ or the $HISQ$ action where the effects of flavor
symmetry breaking are much smaller. It is expected that all actions will give the same result in the continuum
limit resulting in a transition temperature of $(150-160)$MeV. Discrertization effects are also visible for
the renormalized Polyakov loop in the low temperature region for $p4$ and $asqtad$ calculations with $N_{\tau}=8$.
However, the $N_{\tau}=12$ $asqtad$, $HISQ$ and $stout$ data
are all agree. Thus, discretization errors are under control for this quantity.

\section{Correlation functions of static quark anti-quark pair}

One of the most prominent feature of the quark gluon plasma is the presence of chromoelectric (Debye) screening.
The easiest way to study chromoelectric screening is to calculate 
the singlet free energy of static quark anti-quark pair (for recent
reviews on this see Ref. \cite{mehard04,qgp09}),  which is expressed in terms of 
correlation function of temporal Wilson lines in Coulomb gauge
\begin{equation}
\exp(-F_1(r,T)/T)=\frac{1}{N} {\rm Tr} \langle W(r) W^{\dagger}(0) \rangle.
\end{equation}
$L={\rm Tr}  W$ is the Polyakov loop.
Instead of using the Coulomb gauge the singlet free energy can be defined in gauge invariant manner by
inserting a spatial gauge connection between the two Wilson lines. Using such definition the singlet free energy 
has been calculated in $SU(2)$ gauge theory \cite{baza08}.  
It has been found that the singlet free energy calculated this way is close to the result obtained in
Coulomb gauge \cite{baza08}. 
The singlet free energy turned out to be  useful to study quarkonia binding at high temperatures in potential models 
(see e.g. Refs. \cite{mocsy1,mocsy2,mocsy3,mocsy4,mocsy5}).  The singlet free energy also appears naturally in the perturbative
calculations of the Polyakov loop correlators at short distances \cite{plc_new}.

\begin{figure}
\includegraphics[width=7.5cm]{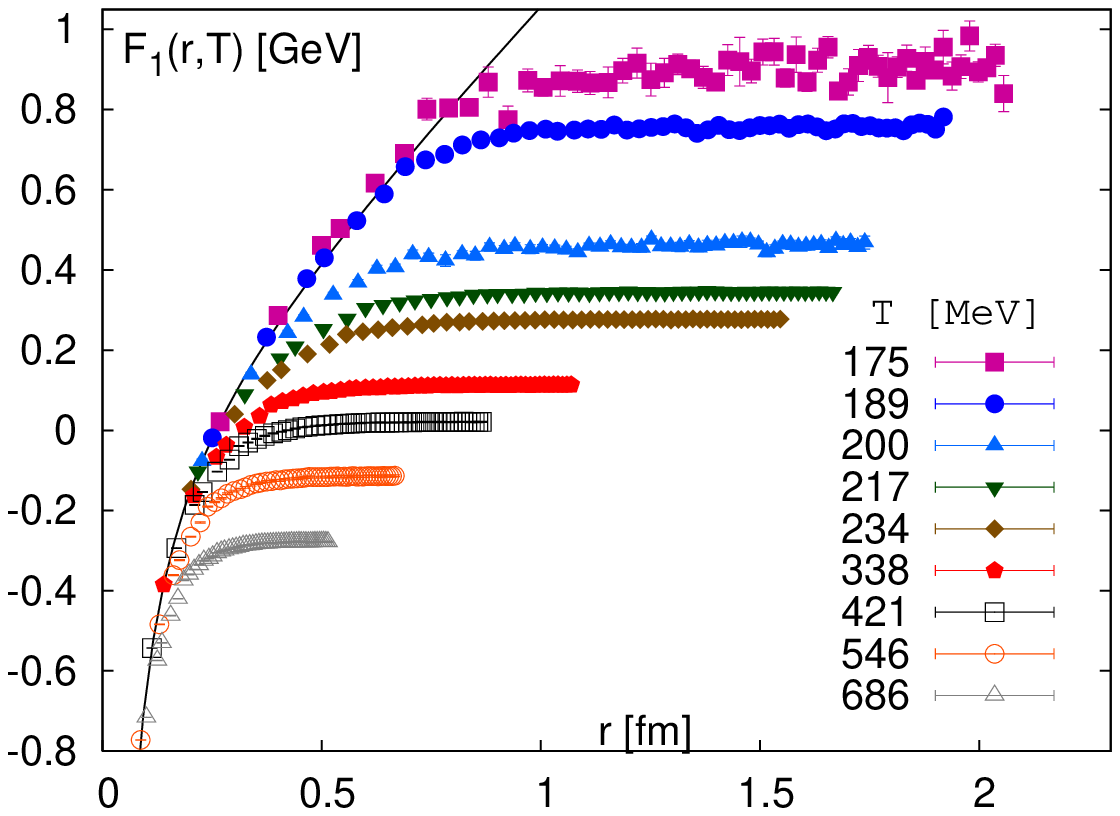}
\includegraphics[width=7.5cm]{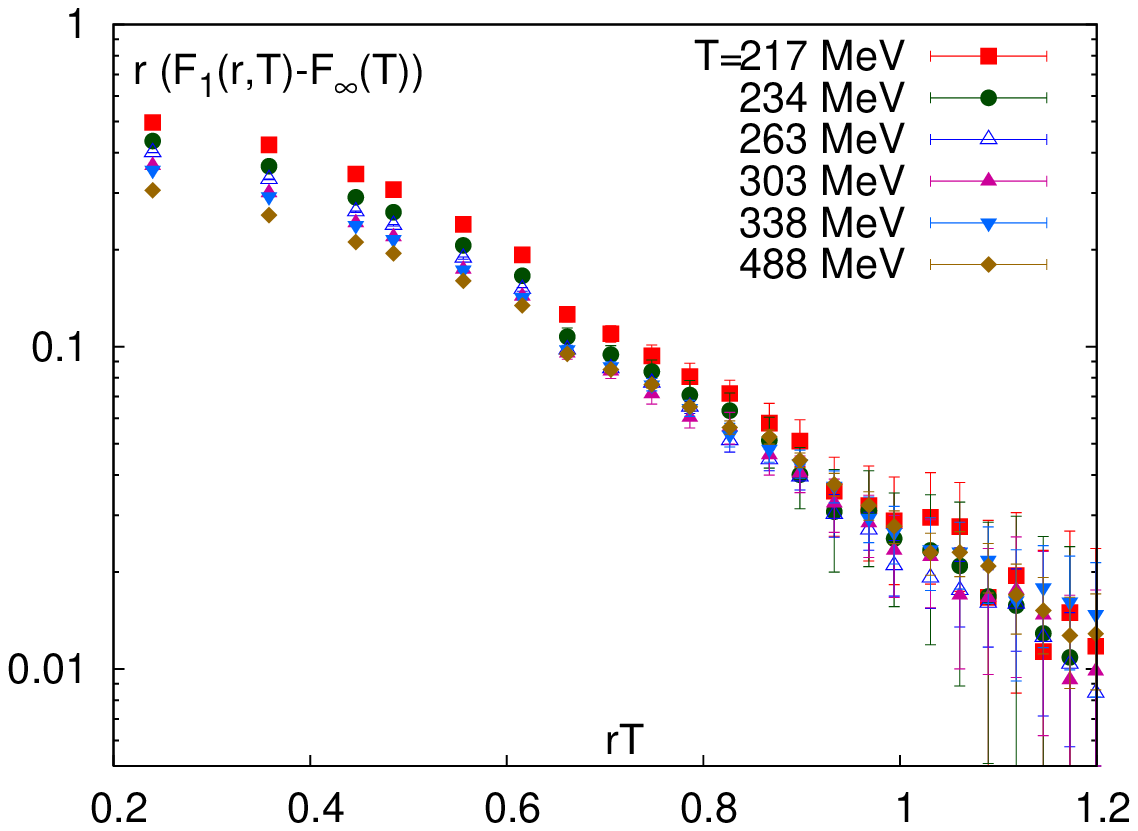}
\caption{The singlet free energy $F_1(r,T)$ calculated in Coulomb gauge on $16^3 \times 4$ lattices (left)
and the combination $F_1(r,T)-F_{\infty}(T)$ as function of $r T$ (right). The solid black line
is the parametrization of the zero temperature potential.}
\label{fig:f1} 
\end{figure}

The  singlet free energy was recently calculated in QCD with one strange quark and two light
quarks with masses corresponding to pion mass of $220$MeV on $16^3 \times 4$ lattices \cite{rbc_f1}. 
The numerical results are shown in Fig. \ref{fig:f1}.
At short distances the singlet free energy is temperature independent and coincides with the 
zero temperature
potential.  In purely gluonic theory the free energy grows linearly with the separation between 
the heavy quark and 
anti-quark in the confined phase. In presence of dynamical quarks the free energy is saturated 
at some finite value 
at distances of about $1$ fm due to string breaking \cite{mehard04,kostya1,okacz05}. 
This is also seen in Fig. \ref{fig:f1}. 
Above the deconfinement temperature the singlet free
energy is exponentially screened at sufficiently large 
distances \cite{okacz02,digal03} with the screening mass proportional to
the temperature , i.e.
\begin{equation}
F_1(r,T)=F_{\infty}(T)-\frac{4}{3}\frac{g^2(T)}{4 \pi r} \exp(-m_D(T) r), ~m_D \sim T.
\end{equation}
Therefore in Fig. \ref{fig:f1} we also show the combination $F_1(r,T)-F_{\infty}(T)$ 
as a function of $r T$. As one 
can see from the figure this function shows an exponential fall-off 
at distances $r T>0.8$. The fact that the slope is the same for all temperatures means
that $m_D \sim T$, as expected.

\section{Temporal meson correlars and spectral functions}
Information on hadron properties at finite temperature as well as the transport coefficients 
are encoded in different spectral functions.
In particular the fate of different quarkonium states in QGP
can studied by calculating the corresponding quarkonium spectral functions 
(for a recent review see Ref. \cite{qgp09}).
On the lattice we can calculate correlation function in Euclidean time. 
This is related to the spectral function via integral
relation
\begin{equation}
G(\tau, T) = \int_0^{\infty} d \omega
\sigma(\omega,T) K(\tau,\omega,T) ,~~
K(\tau,\omega,T) = \frac{\cosh(\omega(\tau-1/2
T))}{\sinh(\omega/2 T)}.
\label{eq.kernel}
\end{equation}
Given the data on the Euclidean meson correlator $G(\tau, T)$ the meson spectral function 
can be calculated
using the Maximum Entropy Method (MEM)  \cite{mem}. For charmonium this was done by 
using correlators calculated on
isotropic lattices \cite{datta02,datta04} as well as  
anisotropic lattices \cite{umeda02,asakawa04,jako07} in the quenched approximation.
It has been found that quarkonium correlation function in Euclidean time show only very small temperature
dependence \cite{datta04,jako07}. In other channels, namely the vector, scalar and axial-vector channels 
stronger temperature dependence was found \cite{datta04,jako07}.
The spectral functions in the pseudo-scalar and vector channels reconstructed from MEM show peak structures which may
be interpreted as a ground state peak \cite{datta04,umeda02,asakawa04}. Together with the weak temperature dependence
of the correlation functions this was taken as strong indication that the 1S charmonia ($\eta_c$ and $J/\psi$) survive
in the deconfined phase to temperatures as high as $1.6T_c$ \cite{datta04,umeda02,asakawa04}. A detailed study of
the systematic effects show, however, that the reconstruction of the charmonium spectral function is not reliable
at high temperatures \cite{jako07}, in particular the presence of peaks corresponding to bound states cannot be
reliably established. Presence of large cutoff effects at high frequencies also complicates the analysis \cite{freelat}.
The only statement that can be made is that the 
spectral function does not show significant changes 
within the errors of the calculations. Recently quarkonium spectral functions have been studied using potential models
and lattice data for the singlet free energy of static quark anti-quark pair \cite{mocsy3,mocsy4,mocsy5}. These calculations show that all
charmonium states are dissolved  at temperatures smaller than $1.2T_c$, but the Euclidean correlators do not show
significant changes and are in fairly good agreement with available lattice data  both 
for charmonium \cite{datta04,jako07} and bottomonium \cite{jako07,dattapanic05}. 
This is due to the fact that even in absence of bound states quarkonium spectral functions
show significant enhancement in the threshold region \cite{mocsy2}.  Therefore previous statements about quarkonia
survival at high temperatures have to be revisited. Exploratory calculations of the charmonium correlators and
spectral functions in 2-flavor QCD have been reported in Ref. \cite{aarts07} and the qualitative behavior
of the correlation functions was found to be similar.

The large enhancement of the quarkonium correlators above deconfinement in the scalar and axial-vector
channel can be understood in terms of the zero mode contribution \cite{mocsy2,umeda07} 
and not due to the dissolution of the $1P$ states as previously thought. 
Similar, though smaller in magnitude, enhancement of quarkonium correlators due to zero mode 
is seen also in the vector channel \cite{jako07}. Here it is related to heavy quark transport \cite{mocsy1,derek}.
Due to the heavy quark mass the Euclidean correlators for heavy quarkonium can be decomposed into
a high and low energy part $G(\tau,T)=G_{\rm low}(\tau,T)+G_{\rm high}(\tau,T)$
The area under the  peak in the spectral functions at zero energy $\omega \simeq 0$ giving the zero mode contribution
to the Euclidean correlator is proportional to some susceptibility, $G^i_{low}(\tau,T) \simeq T \chi^i(T)$, which
have been calculated on the lattice in Ref. \cite{me_hq08}. It is natural to ask whether 
the generalized susceptibilities can be described by a quasi-particle model. The generalized
susceptibilities have been calculated in Ref. \cite{aarts05} in the free theory.
Replacing the bare quark mass entering in  the expression of the generalized susceptibilities by an 
effective temperature dependent masses one can describe the zero mode contribution very 
well in all channels \cite{me_hq08}. 

While temporal correlators are not sensitive to the change in the spectral functions spatial quarkonium
correlation functions could be more sensitive to this. Recent lattice calculations show indication for
significant change in spatial charmonium correlators above deconfinement\cite{mukher}.

The spectral function for light mesons has been calculated on the lattice
in quenched approximation \cite{karsch02,asakawaqm02,aarts_el}. However, unlike in the quarkonia case 
the systematic errors in these calculations are not well understood. 

\section{Conclusions}

In recent years significant progress has been achieved in studying strongly interacting matter
at high temperatures using lattice QCD. Equation of state and transition temperature have been
calculated at several lattice spacings allowing for controlled continuum extrapolations.
Current lattice data suggest a transition temperature for chiral symmetry restoration of $(150-160)$ MeV.
Fluctuations of conserved charges and spatial correlation functions provide detailed
picture of the high temperature medium, namely the nature of the relevant degrees of freedom
(partonic or hadronic) and color screening. Temporal meson correlation functions have been also studied
in lattice QCD but the numerical data are not precise enough to provide detailed information of
the corresponding meson spectral functions. However, these lattice data are useful to constrain the
model calculations of the meson spectral functions.

\section*{Acknowledgments} 
This work was supported by U.S. Department of Energy under
Contract No. DE-AC02-98CH10886.  Computations have been performed using
the USQCD resources on clusters at Fermilab and JLab and the QCDOC supercomputers
at BNL, as well as the BlueGene/L at the New York Center for Computational Sciences (NYCCS).

\end{document}